\documentclass[pdflatex,sn-mathphys-num]{sn-jnl}

\usepackage{epsfig}
\usepackage{subcaption}
\usepackage{calc}
\usepackage{amssymb}
\usepackage{amstext}
\usepackage{amsmath}
\usepackage{amsthm}
\usepackage{multicol} 
\usepackage{url}
\usepackage[most]{tcolorbox}  

\usepackage[table]{xcolor}
\definecolor{maroon}{cmyk}{0,0.87,0.68,0.32}

\usepackage{amssymb}
\usepackage[figuresright]{rotating}

\usepackage{listings} 

\definecolor{codegreen}{rgb}{0,0.6,0}
\definecolor{codegray}{rgb}{0.5,0.5,0.5}
\definecolor{codepurple}{rgb}{0.58,0,0.82}
\definecolor{backcolour}{rgb}{0.95,0.95,0.92}

\lstdefinestyle{mystyle}{
    backgroundcolor=\color{backcolour},   
    commentstyle=\color{codegreen},
    keywordstyle=\color{magenta},
    numberstyle=\tiny\color{codegray},
    stringstyle=\color{blue},
    basicstyle=\ttfamily\footnotesize,
    breakatwhitespace=false,         
    breaklines=true,                 
    captionpos=b,                    
    keepspaces=true,                 
    numbers=left,                    
    numbersep=5pt,                  
    showspaces=false,                
    showstringspaces=false,
    showtabs=false,                  
    tabsize=2
}

\begin{document}

\definecolor{main}{HTML}{5989cf}    
\definecolor{sub}{HTML}{cde4ff}     
\newtcolorbox{boxE}{
    enhanced, 
    boxrule = 0pt, 
    borderline = {0.75pt}{0pt}{main}, 
    borderline = {0.75pt}{2pt}{sub} 
}

\title{Scrum Sprint Planning: LLM-based and algorithmic solutions}

\author{Yuwon Yoon, Kevin Iwan, Madeleine Zwart, Xiaohan Qin, Hina Lee, Maria Spichkova}

\affil{ \orgname{RMIT University}, \orgaddress{ \city{Melbourne}, \country{Australia}}}

\abstract{
Planning for an upcoming project iteration (sprint) is one of the key activities in Scrum planning. 
In this paper, we present our work in progress on exploring the applicability of Large Language Models (LLMs) for solving this problem. 
We conducted case studies with manually created data sets to investigate the applicability of OpenAI’s models for supporting 
 the sprint planning activities.  
 In our experiments, we 
 {applied} three models provided OpenAI: 
  GPT-3.5 Turbo, GPT-4.0 Turbo, and Val. 
  The experiments demonstrated that the results produced by the models aren't of acceptable quality for direct use in Scrum projects. 
}
  
\maketitle
\section{Introduction}
The adoption of Agile software development methodology continues to grow over the last decade~\cite{al2020agile}. 
Scrum methodology~\cite{schwaber2011scrum} has become increasingly popular in managing software development projects. A  survey conducted in 2023 by digital.ai~\cite{Agile17} revealed that more than 2/3 of the participating companies incorporate Agile practices within their software development lifecycle.  

Planning for an upcoming project iteration (sprint) is one of the key activities in Scrum planning. 
However, many junior software developers struggle with planning sprints correctly, if planning is conducted without a close supervision of more experienced developers. The same situation is typically observed by students learning the Agile/Scrum concepts. . 
Given the recent advances of Large Language Models (LLMs) in generating textual outputs across a range of complex tasks, exploring their potential applicability to sprint planning processes appears to be a promising direction. 
\emph{Contributions:}  
In this paper, we present our work in progress on exploring automation strategies, including an analysis of the applicability of LLMs for solving this problem. We aimed to provide an answer to the following research question:

\emph{RQ: What is the current capability of OpenAI LLMs for Scrum sprint planning based on the provided product backlog data and the project description?}

We conducted case studies with manually created data sets to investigate the applicability of OpenAI’s models for supporting 
 the sprint planning activities.  
 In our experiments, we 
 {applied} three models provided OpenAI: 
  GPT-3.5 Turbo, GPT-4.0 Turbo, and Val. 
  The experiments demonstrated that the results produced by the models aren't of acceptable quality for direct use in Scrum projects.  
  The lessons learned presented in this paper can also be applied for teaching Agile/Scrum concepts, aligning with the ideas on learning and teaching presented in~\cite{spichkova2025agile}. Moreover, these lessons can also be applied within project-based learning of Software Engineering, enhancing the structure proposed in~\cite{spichkova2019industry}.

 
\section{Background: Scrum Sprint Planning}
\label{sec:background}

Functional requirements are presented in Agile/Scrum as so-called \emph{Product Backlog} (PB). For all Product Backlog Items (PBIs), the team should specify corresponding priorities and efforts. 

In Scrum, the set of priorities to select from is typically specified as \emph{Critical} (should be implemented urgently), \emph{High} (to be implemented as soon as possible), \emph{Medium} (important, but can be implemented in the second half of the project), and \emph{Low} (nice-to-have functionality). Defining priorities should be based on customers' and the product owner's decisions. Items in the PB should be sorted by their priority, having higher priority items on top, which should simplify the selection of items for an upcoming sprint. 

Efforts are specified using so-called \emph{story points}, using Fibonacci numbers. 

A set of the PBIs selected for a sprint forms a \emph{Sprint Backlog}.  
In the case that all priorities, efforts, and dependencies among PBIs are specified correctly, sprint planning should be straightforward: the team should select from the top of the PB a set of the highest-priority items (which haven't been completed yet), taking into account efforts associated with PBIs and dependencies. 

The Scrum routine typically consists of several meeting types:
\begin{itemize}
    \item 
    Sprint planning meetings: These meetings should be the very first activity within the sprint, where the upcoming sprint is planned and the corresponding sprint backlog is specified.
    \item  
    Daily Scrum meetings: These daily meetings are typically very short (10-15 minutes) and aim to update all team members on everyone's progress and potential issues.
    \item 
    Product backlog refinement meeting, where the updates and refinements of the product backlog have to be discussed.
    \item
    Sprint review meetings are held at the end of the sprint to demonstrate the sprint results to the Product Owner and customers. This should be the pre-last activity in the sprint;
    \item 
    Sprint retro meetings (retrospectives) should be the very last activity within the sprint, where the team meets internally (without the Product Owner and customers) to discuss their progress within the concluded sprint, to reflect on what went well and what went not so well, as well as to analyse the lessons learned and plan corresponding action items to improve their process.
\end{itemize}

\section{Related Work}
\label{sec:related}

\subsection{Sprint planning in Agile/Scrum methodologies}

Numerous studies aim to enhance sprint planning. For instance, a decision support system for sprint planning in Scrum is proposed in~\cite{alhazmi2018decision}. Additionally, an optimization model for multi-sprint planning is introduced in~\cite{golfarelli2013multi}.
Improvements to the Scrum planning process have been outlined in \cite{app15010202}.
Unlike the works mentioned above, we aim to investigate the applicability of OpenAI's Large Language Models (LLMs) for sprint planning.

\subsection{Requirements prioritisation and effort estimations}

The prioritisation of user stories serves as a crucial foundation for effective sprint planning. 
Incorrectly assigned priorities can negatively affect the overall planning process. 
Numerous studies have addressed the prioritisation of product backlog items within Agile/Scrum settings, as well as requirements prioritisation for traditional methodologies, see for example~\cite{al2013prioritizing}, \cite{sachdeva2018prioritizing}, \cite{jarzkebowicz2020agile}, and \cite{borhan2022requirements}. 
In our study, assume that the priorities have already been assigned to the user stories.

There are also many research studies that focus on effort estimation, see, for example, the works presented in 
\cite{zahraoui2015adjusting}, \cite{alostad2017fuzzy}, \cite{arora2020systematic}, \cite{govil2022estimation}, \cite{butt2022software}.
While effort estimation serves as an essential foundation for planning, it falls outside the scope of our current project. 
In our framework, we proceed with the assumption that the effort estimation has already been conducted for the analysed product backlog.

\subsection{AI-based approaches for Scrum process}

A comprehensive literature review on the use of LLMs in Software Engineering (SE) has been introduced in~\cite{hou2024large}. This review covers the analysis of 395 research articles published between 2017 and 2024, categorising the various LLMs applied to SE tasks, reviewing the methodologies used, and examining optimisation and evaluation strategies. Another survey on this topic, presented in \cite{fan2023large}, places particular emphasis on the open research challenges associated with applying LLMs in the SE domain.

Another literature review focused on the perspectives of junior software developers regarding the adoption of LLMs, see~\cite{ferino2025junior}. 
The use of speech recognition tools for enhancing retrospective analysis has been analsed in the study presented in~\cite{gaikwad2019voice}. The study utilised Google Home and Amazon Alexa tools to investigate potential improvements in the timeboxing of a retrospective by using voice-activated commands.

A case study exploring ChatGPT's potential to support requirements elicitation has been presented in \cite{ronanki2023investigating}. An approach for generating program specifications using LLMs was introduced in \cite{ma2024specgen}, while \cite{oswal2024transforming} proposed a vision for transforming traditional functional requirements into user stories using the GPT-3.5 model.

In our previous work~\cite{ENASE25Spichkova}, we conducted an  analysis on whether LLMs might be applicable for support of another type of Scrum meetings, the Retrospective (retro) meetings. We aimed to investigate whether an AI-based solution might help to simplify these meetings and to make them safer from the psychological side. We also introduced a research prototype, RetroAI++, which aim is to automate and streamline Agile/Scrum routine within Sprint Planning and Retrospectives~\cite{spichkova2025advanced}.

In~\cite{cabrero2024exploring}, the authors analyse the applicability of AI assistants for another type of Daily Scrum meetings. The results of the study highlighted that AI cannot replace Scrum Masters, but can be potentially useful to generate contextualised insights. 

\subsection{Research embedded in teaching and learning process}

This study has been conducted within the employability initiative 
proposed at the RMIT University, see \cite{spichkova2017autonomous,simic2016enhancing}.
To foster the curiosity of Bachelor's and Master's students regarding research in Software Engineering, we propose the incorporation of research and analysis components into the projects as an optional bonus task. The research projects were typically short, corresponding to approx. one week of a full-time study. They have been typically conducted by teams of four or five students, sponsored by industrial partners. The topic of the research projects aligned with the topics related to the software development project to be conducted within the semester, which led to a broad range of topics to be covered by several teams.
For example, Sun et al.~\cite{sun2018software} analysed Human-Computer interaction aspects related to an autonomous humanoid robot, whose task is to guide visitors through a VXlab. George et al.~\cite{george2020usage} conducted a study on the visualisation of the usage of the AWS service. 
Zhao et al.~\cite{zhao2025visualisation} also worked on the visualisation aspects, but their study focused on the CIS benchmark scanning results.

\section{{Case study: Methods}}
\label{sec:methodology}

LLMs have the potential to enhance productivity by assisting with the execution of repetitive and time-consuming tasks. 
However, in some cases, they don't provide any significant benefit wrt. algorithmic solutions.  
Our study mainly focuses on the applicability OpenAI’s ChatGPT for facilitating the sprint planning activities.  
 In our experiments, we applied three models provided OpenAI: 
  GPT-3.5 Turbo, GPT-4.0 Turbo, and Val. 

 As the data set, we applied product backlogs and requirements elicitation notes from simulated Scrum projects: Baking4U  and Online Patient Management projects.
 These product backlogs and notes have been used for learning and teaching purposes at RMIT University within a course ISYS1108/3474 Software Engineering Project Management. The course mainly focuses on Agile/Scrum project management. 
The requirements elicitation notes provide the basis for the prioritisation of  PBIs. These notes are presented as a 1-page PDF document for each project. The document contains a general project description as well as information items' priorities.  

The complete PBs have been manually created by the last author based on the elicitation notes. 
The PB for the Baking4U project consisted of 25 items, while the 
PB for the  Online Patient Management project consisted of 22 items. 
These PBs covered all typical fields including IDs, user stories, priorities, efforts, and status.  The PBs have been used for marking purposes in the corresponding course assignments in 2023 and 2024. Within our case studies, these PBs have been used  
\begin{itemize}
    \item as a benchmark solution to analyse the correctness of LLM responses (the complete backlogs, including priorities), and
    \item
    as input data for prioritisation tasks (backlog items have been limited to user stories, IDs, estimation and status, i.e., the priority field has been removed).
\end{itemize}
The PB (input) data have been provided to the ChatGPT models in the form of JSON data set, see below a sample excerpt presenting a single PBI:
 
{\footnotesize{
\begin{verbatim}
 {  "shortId": 19,
    "title": "As a staff user, I want to add,  update, or remove bread and pastry listings, 
    so that our customers always have access to our most current offerings and accurate 
    information about our products.",
    "estimation": 5,
    "status": "To Do"
    },
\end{verbatim}
}}

\section{{Case study: Prompt Engineering}} 
\label{sec:prompts}

The objective of prompt engineering is to optimise inputs for LLMs in order to enhance the quality of their output performance, see~\cite{white2023prompt}. 
The strategy for prompt engineering in our study was to specify instructions as clearly as possible and to add more explicit constraints to the input if the outputs demonstrated that the model might require the corresponding information to improve the results. 
In this work, we limited the overall set to selected prompts, which resulted in more consistent and accurate results. 
to demonstrate the process of elaboration as well as the lessons learned.  


\begin{boxE}
\footnotesize{
\textbf{Prompt 1:}\\
What order should the following stories be completed in?\\ Order the items in each list based on the item's dependencies. The order should align with the project description.\\ 
Project Description: Refer to Appendix A.\\
Output Format Requirement: The models were required to return 3 lists (high, medium, and low priority) with task short IDs formatted as: ['shortId1', 'shortId2', 'shortId3'...]
}
\end{boxE}

Applying Prompt 1, we observed many inconsistencies in the results, including inconsistencies among sessions,  which might be related to the lack of context that the project will be conducted using Scrum. Therefore, we added the corresponding information to the next prompt as \emph{This is a SCRUM project, so please follow SCRUM principles when determining the order of completion.}

Adding these instructions %
did not noticeably improve the consistency of the task order. 
This issue might be caused by the lack of 
background knowledge on Scrum, therefore, we expanded the next prompt by adding core Scrum principles, such as
In the visualisation of the prompts below, we replace by $<...>$ all content repeated from Prompt 2. While working with the ChatGPT models, the complete prompts have been provided to the models.

 \begin{boxE}
 \footnotesize{
 \textbf{Prompt 3:}\\
 $<...>$ \\
\emph{Ensure that tasks are ordered based on their dependencies, which should be inferred from the project description, card titles, and descriptions. Tasks that are dependencies for other tasks should be completed first.	\\ 
- The project description provides context for the overall goals and milestones. Tasks that align with these goals or are critical to achieving milestones should be prioritised higher.	 \\
- Use the card titles to quickly identify the main functionality or feature the task addresses. Prioritise tasks that implement core features identified in the titles.	 \\
- The card descriptions should be used to uncover details about dependencies, complexity, and potential impact on the project’s success. Use these insights to refine the task order.	 \\
- Consider the status of each task when determining the order. Tasks that are 'Ready to Present' or 'In Progress' should generally be completed before tasks that are still in the 'To Do' phase.	 \\
- Sort tasks primarily by their priority level (High, Medium, Low). Within each priority level, sort tasks by their dependencies and readiness status.\\}
**Formatting Requirement**: The output **must** strictly adhere to the following format:\\
 - The output should consist of three lists, labelled HIGH, MEDIUM, and LOW, in that order.\\
 - Each list should only include shortId values, enclosed in square brackets, and formatted as: ['shortId1', 'shortId2', 'shortId3', ...].	 \\
- Ensure the lists are correctly ordered and labelled, with no additional text or formatting.
}
\end{boxE}

 Providing more explicit instructions did not reduce the inconsistencies across sessions. The model's variability in interpreting task dependencies and priorities persisted. 

\begin{boxE}
\footnotesize{
\textbf{Prompt 4:}\\
 Given the following project description and lists of tasks, order the tasks in each list based on their dependencies, ensuring the order aligns with the project's goals and SCRUM principles.\\
Project Description: Refer to Appendix A.\\
Instructions:
Dependency-Based Ordering: Order the tasks based on their dependencies. Tasks that are prerequisites for other tasks should be completed first.\\
Project Goals Alignment: Consider how each task aligns with the overall goals and milestones of the project. Prioritise tasks that are critical to achieving these goals.
Task Titles \& Descriptions: Use the titles and descriptions to quickly identify the main functionality or feature of each task. Prioritise tasks that implement foundational or core features.\\
Task Status: Consider the status of each task. Tasks that are "Ready to Present" or "In Progress" should generally be completed before those that are still in the "To Do" phase.
Final Sorting: After considering dependencies, readiness status, and project goals, finalise the order within each priority level.\\
Output Format:
Ordered Lists: Return the ordered tasks as three separate lists for high, medium, and low priority tasks.
Formatting: Each list should only include shortId values, formatted as: ['shortId1', 'shortId2', 'shortId3', ...]\\
Explanation:
Dependencies Identified: Provide a thorough explanation of the dependencies among all tasks and how they influenced the final ordering. Ensure that the reasoning includes every task, explaining how each was prioritised or sequenced in relation to the others.
}
\end{boxE}

\section{{Case studies: Results}} 
\label{sec:results}

In this section, we provide a summary of the primary findings and briefly discuss solutions we propose to solve the observed issues while applying ChatGPT models to support the planning of sprints.  

In our analysis,  
    $N$ denotes the set size (the number of items in a backlog), 
    $N_C$ denotes the number of items whose priorities have been identified correctly, 
    $P_C$ denotes the percentage of items whose priorities have been identified correctly, i.e., $P_C = N_C / N$.

\textbf{Case Study 1:} The experiments have been conducted using the Baking4U data set and Prompt 4 presented in the previous section. The overview of the quantitive analysis of case study results ($P_C$ values) is presented in Table~\ref{tab:analysis}.

\begin{table}[ht!]
    \centering
    \small{ 
    \begin{tabular}{l|l l l}
         & \multicolumn{3}{c}{$P_C$, \%}\\
         & GPT-3.5  & GPT-4.0  & Val \\
         \hline      \hline
       $\mu$ (mean) & 55 & 59 & 71 \\
       $\sigma$ (standard deviation) & 8 & 15 & 12\\
       \hline
        \textbf{E1, average}     & \textbf{48}   & \textbf{56}  & \textbf{84}   \\
        \textbf{E2, average} & \textbf{57}& \textbf{62} & \textbf{57} \\
        E2, S1 & 72 & 92 & 60 \\
        E2, S2 & 48 & 44 &  52 \\
        E2, S3 & 56 & 56 & 68 \\
        E2, S4 & 52 & 56 &  56 \\
        E2, S5 & 56 & 60&  48 \\
    \end{tabular}}
    \caption{Quantitive analysis of Baking4U case study results }
    \label{tab:analysis}
\end{table}

In Experiment 1 (E1), 
the format of the data set was like presented in Section~\ref{sec:methodology}, 
while in Experiment~2 (E2), 
we limited the data set to user stories and IDs (i.e., the fields \emph{estimation} and \emph{status} have been removed) and conducted 5 sessions with each model (S1, \dots, S5).  The removal of effort and status from the task prioritisation process did not significantly improve the consistency or correctness of the models' performance. Despite focusing solely on task dependencies, all models still exhibited considerable variability and inconsistency across multiple sessions. This indicates that even without the influence of effort or status, the models struggled to maintain stability.

The highest mean value of $P_C$ was observed by Val (71\%), while the lowest was observed by GPT-3.5 (55\%). 
These results suggest that all three models don't provide good enough prioritisation results, so that using the capacity of these LLMs without any additional training and fine-tuning is not advisable. 

Moreover, all three models demonstrated showed significant variability across sessions, with $P_C$ values fluctuating for GPT-3.5 Turbo between 48\% and 72\%, while for GPT-4.0 Turbo they were fluctuating between  44\% and 92\%, and for Val the values were between 52\% and 84\%. This suggests that the model's task prioritisation accuracy is inconsistent. 
The wide range in errors highlights the models' challenges in providing stable prioritisation.

\textbf{Case Study 2:} The experiments have been conducted using the OPM data set. Similarly to Experiment 2 of Case Study 1,  
we limited here the data set to user stories and IDs (i.e., the fields \emph{estimation} and \emph{status} have been removed), as these settings provided better results for majority of the models. 
For the OPM data set, all models demonstrated better consistency between models and across sessions. The reason for this might be that the user stories in this data set were simpler/shorter. 
Nevertheless, the results obtained within this case study were similar to the results of Case Study~1.

The overall correctness of prioritisation wasn't high enough to consider any of three models for direct use.
{GPT-3.5} model demonstrated lowest results, with significant variability among sessions in the classification of high and medium priority items. GPT-4.0 generally performed better, consistently demonstrated correct prioritisation of some items. Val was the most consistent model in correctly identifying all high priority items, while having mistakes in prioritisation of medium and low priority items.
%

\section{{Limitations and Threats to validity}}
\label{sec:limitations}

Our experiments face several threats to validity. This section identifies, categorizes, and discusses our attempts to mitigate them.

\emph{External threats to validity}: 
The first threat is the size of the dataset, which was limited to two product backlogs, both contain a relatively small set of user stories.  
However, this scale of product backlogs is reasonable for small-scale projects. 
The second threat is the limited number of runs for each prompt presented in the experiment analysis in Sections~\ref{sec:methodology} and~\ref{sec:results}. 
During our experiments, we noticed that the results of prompt executions may differ slightly. In other words, running the same prompt multiple times might yield responses from ChatGPT that are not exactly the same. 

\emph{Internal threats to validity}: 
%
The classification and analysis of the ChatGPT responses were conducted manually by the authors. To address the issues of inconsistent conclusions, a peer-review strategy was employed within the project.

\section{{Conclusions and Future Work}}
\label{sec:conclusions}

In this paper, we outline our continued research aimed at enhancing Agile/Scrum processes through the integration of GenAI approaches.  We discussed our experiments with OpenAI's ChatGPT-4 GPT-3.5 Turbo, GPT-4.0 Turbo, and Val to analyse the applicability of these models for sprint planning and summarised the core lessons learned from these experiments. The result of our study demonstrates that algorithmic solutions might be preferable to support junior software developers and students mastering Agile/Scrum concepts related to sprint planning.  
%

\section*{{Acknowledgements}}

We would like to thank Shine Solutions for sponsoring this project under the research grant PRJ00002505, and especially Branko Minic and Adrian Zielonka for sharing their industry-based expertise and advice.

 
 

\end{document}